\documentclass{llncs} 
\usepackage{graphicx}
\usepackage{amssymb,amsmath,mathtools}
\usepackage{enumerate}
\usepackage{tikz}
\usepackage{url}

\usepackage{booktabs}
\newcommand{\setmeter}[2]{\ensuremath{%
\vcenter{\offinterlineskip
\halign{\hfil##\hfil\cr
$\scriptstyle#1$\cr
\noalign{\vskip1pt}
$\scriptstyle#2$\cr}
}}%
}

\newcommand{\comment}[2]{\textcolor{#1}{#2}}

\newcommand{\cc}[1]{\comment{black}{#1}}
\newcommand{\mg}[1]{\comment{black}{#1}}

\title{What were you expecting? Using Expectancy Features to Predict Expressive Performances of Classical Piano Music}
\author{Carlos Cancino-Chac\'on\inst{1,2}\and Maarten Grachten\inst{2} \and David R.~W.~Sears\inst{2}\and Gerhard Widmer\inst{1,2}}
\institute{Austrian Research Institute for Artificial Intelligence\\
\email{carlos.cancino@ofai.at}
\and
Department of Computational Perception, Johannes Kepler University Linz
\email{\{maarten.grachten,david.sears,gerhard.widmer\}@jku.at}
}
\titlerunning{Using Expectancy Features to Predict Expressive Performances}
\authorrunning{Cancino et al.}
\begin{document}
\maketitle

\begin{abstract}
In this paper we present preliminary work examining the relationship between the formation of expectations and the realization of musical performances, paying particular attention to expressive tempo and dynamics. To compute features that reflect what a listener is expecting to hear, we employ a computational model of auditory expectation called the \textit{Information Dynamics of Music} model (IDyOM). We then explore how well these expectancy features -- when combined with score descriptors using the Basis-Function modeling approach -- can predict expressive tempo and dynamics in a dataset of Mozart piano sonata performances.
Our results suggest that using expectancy features significantly improves the predictions for tempo.
\vspace{-0.3cm}
\end{abstract}
\section{Introduction}\label{sec:introduction}
\vskip-1ex
Computational models of musical expression can be used to explain the way certain properties of a musical score relate to an expressive rendering of the music~{\cite{Widmer:2004bh}.
However, existing models tend to use  a combination of high- and low-level \emph{hand-crafted features} reflecting structural aspects of the score that might not necessarily serve as perceptually relevant features.
An example of such a model is the Basis-Function modeling approach (BM) \cite{Grachten:2012hk}.

To examine the relationship between the formation of expectations during music listening on the one hand, and the realization of musical performances on the other, Gingras et al.~\cite{Gingras:2016bm} employed the \textit{Information Dynamics of Music} model (or IDyOM)~\cite{Pearce:2005th}, a probabilistic model of auditory expectation that computes information-theoretic features relating to the prediction of future events.
In their study, these information-theoretic features were shown to correspond closely with temporal characteristics of the expressive performance, which suggests that the performer attempts to decrease the processing burden on listeners during perception by slowing down at unexpected/uncertain moments and speeding up at expected/certain ones.

Here we present preliminary work to support the claim that expectancy measures can inform predictions of expressive parameters related to tempo and dynamics. 
\cc{We extend the work in~\cite{Gingras:2016bm} in two ways.
First, rather than simply demonstrating that expectancy measures are related to expressive performances, we show that the use of expectancy features improves the predictive quality of models using other score descriptors, thus providing a more comprehensive framework for the modeling of expressive performances in music of the common-practice period.
Second, as opposed to \emph{fitting} the expectancy features to each performance (i.e.~training and testing the model on the same performance), the models presented in this paper are evaluated} \mg{by measuring their prediction error on unseen pieces.}

The rest of this paper is organized as follows:
Section \ref{sec:modeling_expressive_performances} presents our formalization of expressive parameters, describes the score and expectancy features employed in this study, and finally outlines the regression model used to predict the expressive parameters.
Section \ref{sec:experiments} describes the empirical evaluation of the proposed approach, the results of which are discussed in Section \ref{sec:results_and_discussion}. Finally, conclusions are stated in Section~\ref{sec:conclusions}.
\vskip-2ex

\section{Modeling expressive performances}\label{sec:modeling_expressive_performances}
\vskip-1ex
In this section we provide a brief description of the proposed framework.
First we describe how expressive dynamics and tempo are encoded.
Second, we describe the expectancy and score features.
Finally we describe the recurrent neural network (RNN) models used to connect the input features to the expressive targets.

\subsection{Targets: Expressive Parameters}\label{sec:expressive_parameters} 

An \emph{expressive parameter} is a numerical descriptor that corresponds to common concepts involved in expressive piano performance.
We take the local \emph{beat period ratio} $(\mathit{BPR})$ as a proxy for \emph{musical tempo}. 
\cc{We average the performed onset times of all notes occurring at the same score onset and then compute the $\mathit{BPR}$ by taking the slope of the averaged onset times (in seconds) with respect to the score onsets (in beats) and dividing the resulting series by its average beat period.}
For \emph{dynamics}, we treat the performed MIDI velocity as a proxy for loudness.
We take the maximal performed MIDI velocity per score onset, divided by $127$.
This expressive parameter will be denoted $\mathit{VEL}$.
To explore how well the expectancy and score features describe the \emph{relative} changes in $\mathit{BPR}$ and $\mathit{VEL}$, we also calculate their first derivatives, denoted by $\mathit{BPR}_{d}$ and $\mathit{VEL}_{d}$, respectively.
Furthermore, including the derivative time series allows us to compare our findings with the results obtained in \cite{Gingras:2016bm}.

\subsection{Features: Multiple Viewpoints}\label{sec:expectancy_and_score_features}

\subsubsection{Expectancy Features}\label{sec:expectancy_features}
 IDyOM provides a conditional probability distribution of a musical event, given a preceding sequence of events, i.e.~$p(v_n\mid v_{n-1}, v_{n-2}, \dots)$.
Following \cite{Gingras:2016bm}, we use IDyOM to estimate two information-theoretic measures representing musical expectations:
\newcounter{features}
\newcounter{featureGroups}
\begin{enumerate}
\item {\textbf{Information content} ($IC$).}
The $IC$ measures the unexpectedness of a musical event, and is computed as $IC(v_n) = - \log p(v_n \mid v_{n-1},v_{n-2}, \dots)$.

\begin{enumerate}
\item $IC_m$.
The information content for each melody note.
This value is computed using a model that is trained to predict the next chromatic melody pitch using a selection of melodic viewpoints, such as pitch interval (i.e.~the arithmetic difference between two consecutive chromatic pitches, measured in MIDI note values), and contour (whether the chromatic pitch sequence rises, falls or remains the same).
IDyOM performs a stepwise selection procedure that combines viewpoint models if they minimize model uncertainty as measured by corpus cross entropy \cite{Sears:2016ud}.

\item $IC_c$.
Estimation of the $IC$ computed for the combination of pitch events (a proxy for harmony) at each score onset.
IDyOM predicts the next combination of vertical interval classes above the bass \cc{(see Score Features \ref{sec:pitch_vintcc})}.
\setcounter{features}{\value{enumi}}
\end{enumerate}
\item \textbf{Entropy} is a measure of the degree of choice or uncertainty associated with a predicted outcome.
The entropy can be computed as~$H(v_n) = \mathbb{E}\{-\log p(v_n\mid v_{n-1}, v_{n-2}, \dots )\}.$
\begin{enumerate}

\item $H_m$.
Entropy computed for each chromatic pitch in the melody.
\item $H_c$.
Entropy computed for the combined pitch events at each score onset.
\end{enumerate}
\end{enumerate}
\vspace{-0.1cm}
\subsubsection{Score Features}\label{sec:score_features}
Following \cite{Grachten:2012hk}, we include low-level descriptors of the musical score that have been shown to predict characteristics of expressive performance.
\begin{enumerate}
\item \textbf{Pitch.}
\begin{enumerate}
\item $(pitch_h, pitch_l, pitch_m)$.
Three features representing the chromatic pitch (as MIDI note numbers divided by $127$) of the highest note, the lowest note, and the melody note at each onset.
\item $(vic_1, vic_2, vic_3)$.\label{sec:pitch_vintcc}
Three features describing up to three vertical interval classes above the bass, i.e.~the intervals between the notes of a chord and the lowest pitch, excluding pitch class repetition and octaves.
For example, a $C$ major triad ($C$, $E$, $G$), starting at $C_4$ would be represented as $(\begin{array}{cccc}pitch_l &vic_1 &vic_2 &vic_3\end{array}) =  (\begin{array}{cccc}\frac{60}{127}& \frac{4}{11}& \frac{7}{11}&0\end{array})$, \cc{where $0$ denotes the absence of a third interval above $C_4$, i.e.~the absence of a fourth note in the chord.}
\end{enumerate}

\item \textbf{Metrical position.}
\begin{enumerate}
\item $b_{\phi,t}$.
The relative location of an onset within the bar,  computed as $b_{\phi, t} = \frac{t \mod B}{B}$, \cc{where $t$ is the temporal position of the onset measured in beats from the beginning of the score, and $B$ is the length of the bar in beats}.
\item $(b_d, b_s, b_w)$.
Three binary features (taking values in $\{0, 1\}$) encoding the metrical strength of the $t$-th onset.
$b_d$ is nonzero at the downbeat (i.e.~whenever $b_{\phi, t} = 0$);
$b_s$ is nonzero at the secondary strong beat in duple meters (e.g.
quarter-note 3 in \setmeter{4}{4}, and eighth-note 4 in \setmeter{6}{8}), and $b_w$ is nonzero at weak metrical positions (i.e.~whenever $b_d$ and $b_s$ are both zero).
\end{enumerate}
\end{enumerate}

\subsection{Recurrent Neural Networks}\label{sec:regression_models}
\vskip-1ex
RNNs are a state-of-the-art family of neural architectures for modeling sequential data. 
Following \cite{CancinoChacon:2017ht,Grachten:2017ub}, we use bidirectional RNNs as non-linear regression models to assess how well the features described above predict expressive dynamics and tempo.
In this work, we use an architecture with a composite bidirectional hidden layer with 5 units, consisting of a forwards and backwards long short-term memory layer (LSTMs).
\vspace{-0.3cm}

\section{Experiments}\label{sec:experiments}
\vskip-1ex
We perform a 5-fold cross-validation to test the accuracy of the predictions of three models trained on different feature sets for each expressive parameter: a model trained only using expectancy features (\textbf{E}), a model trained only using score features (\textbf{S}), and a model trained on both expectancy and score features (\textbf{E+S}).
Each model is trained/tested on 5 different partitions (folds) of a dataset, which is organized into training and test sets, such that each piece in the corpus occurs exactly once in the test set.

For this study we use the Batik/Mozart corpus, which consists of recordings of 13 Mozart piano sonatas (39 movements) by Austrian pianist Roland Batik performed on a computer controlled B\"osendorfer SE~\cite{Flossmann:2011ww}.
Melody voices were identified and annotated manually in this dataset.
For each fold, we use 80\% of the pieces for training and 20\% for testing.
The parameters of the models are learned by minimizing the mean squared error on the training set\footnote{A repository containing the code is available online: \url{https://github.com/neosatrapahereje/mml2017_expression_expectation}.
}.
We evaluate model accuracy with the coefficient of determination $R^2$ and Pearson's $r$.
\vspace{-0.3cm}

\section{Results and Discussion}\label{sec:results_and_discussion}
\vskip-1ex
The results of the 5-fold cross-validation are shown in Table \ref{tab:cv_results}.
To examine the differences between the $R^2$ values of the \textbf{E}, \textbf{S}, and \textbf{E+S} feature sets we performed a separate one-way ANOVA for each expressive parameter \cc{($BPR$, $BPR_d$, $VEL$ and $VEL_d$)}.
These differences were statistically significant in all cases at the $p<0.05$ level as measured by Fisher's $F$ ratio.
The same trend emerged for most expressive parameters, with \textbf{E+S} outperforming the other models, although post-hoc pairwise comparisons using Tukey's HSD only revealed a significant difference for $\mathit{BPR}_{d}$.
These results therefore suggest that the models including both expectancy and score features better predict expressive tempo \mg{than expressive dynamics}.
\cc{Furthermore, although not directly comparable, the values for $R^2$ and $r$ in Table \ref{tab:cv_results} seem to be on par with those reported on Chopin piano music using the BM approach~\cite{Grachten:2017ub}.}

\cc{The fact that the use of expectancy features improves model performance for expressive tempo but not for dynamics might be due to the relation of expressive tempo to structural properties of the music, such as phrase-final lengthening, such as the final \textit{ritardando} at the end of a piece~\cite{Honing:2006ja}.
Since expectation features also relate to music structure in the sense that music tends to be more unpredictable at boundaries between musical segments than within segments \cite{Pearce:2008tv},
this may in part explain why the models are better at predicting changes in expressive tempo $BPR_d$.}

\setlength{\tabcolsep}{9pt}
\begin{table}[t!]
\centering
\newcommand{\cwa}{\hspace{1em}}
\newcommand{\cwb}{\hspace{2.5em}}
\newcommand{\cwc}{\hspace{4em}}
\begin{tabular}{cl@{\cwa}l@{\cwb}l@{\cwa}l@{\cwc}l@{\cwa}l@{\cwb}l@{\cwa}l}
\toprule
& \multicolumn{4}{c@{\cwc}}{\textbf{Tempo}} & \multicolumn{4}{c}{\textbf{Dynamics}}                                                                     \\
\textbf{Feature} & \multicolumn{2}{c@{\cwb}}{$\mathit{BPR}$} & \multicolumn{2}{c@{\cwc}}{$\mathit{BPR}_{d}$} & \multicolumn{2}{c@{\cwb}}{$\mathit{VEL}$} & \multicolumn{2}{c}{$\mathit{VEL}_{d}$} \\
\textbf{Set} & \multicolumn{1}{c@{\cwa}}{$R^2$} & \multicolumn{1}{c@{\cwb}}{$r$} & \multicolumn{1}{c@{\cwa}}{$R^2$} & \multicolumn{1}{c@{\cwc}}{$r$} & \multicolumn{1}{c@{\cwa}}{$R^2$} & \multicolumn{1}{c@{\cwb}}{$r$} & \multicolumn{1}{c@{\cwa}}{$R^2$} & \multicolumn{1}{c}{$r$} \\ \midrule
\textbf{E}      & 0.038 & 0.201 & 0.067 & 0.259 & 0.234 & 0.496 & 0.185 & 0.429 \\
\textbf{S}      & 0.065 & 0.289 & 0.105 & 0.326 & 0.299 & 0.569 & 0.244 & 0.494 \\
\textbf{E + S}  & 0.072 & 0.288 & 0.124 & 0.351 & 0.312 & 0.574 & 0.230 & 0.477 \\ \bottomrule
\end{tabular}
\vskip1ex
\caption{Predictive results for expressive tempo and dynamics, averaged over all pieces on the Batik/Mozart corpus.
A larger $R^2$ and $r$ means better performance.}
\label{tab:cv_results}
\vspace{-0.5cm}
\end{table}

Figure \ref{fig:sensitivity} shows 2D differential sensitivity maps that examine the contribution of each feature to the output of the model trained on all features (\textbf{E+S}). 
\cc{
Although these plots show that the score features have a more prominent role in predicting expressive tempo, as suggested by the results in Table \ref{tab:cv_results}, we will focus here on the contribution of the expectancy features.
}
On the one hand, the plots suggest a tendency for the performer to slow down if the next melodic events are unexpected or uncertain 
\cc{(see the reddish hue in $H_m$ and $IC_m$ for time-steps $>\tau$ in the right plot)},
and to speed up if the previous melodic events were unexpected or uncertain 
\cc{(the bluish hue in $H_m$ and $IC_m$ for time-steps $<\tau$ in the right plot)},
which is consistent with the findings reported in \cite{Gingras:2016bm}.
On the other hand, while a passage consisting of uncertain harmonic events contributes to an overall slower tempo \cc{(the reddish hue in row $H_c$ in the left plot)}, there is a tendency to speed up if the current harmonic event is unexpected or uncertain \cc{(the bluish hue in $H_c$ and $IC_c$ at $\tau$  in the right plot)}. 


\begin{figure}[t]
\begin{center}
\includegraphics[trim={10 15 15 10}, clip, width=0.49\linewidth]{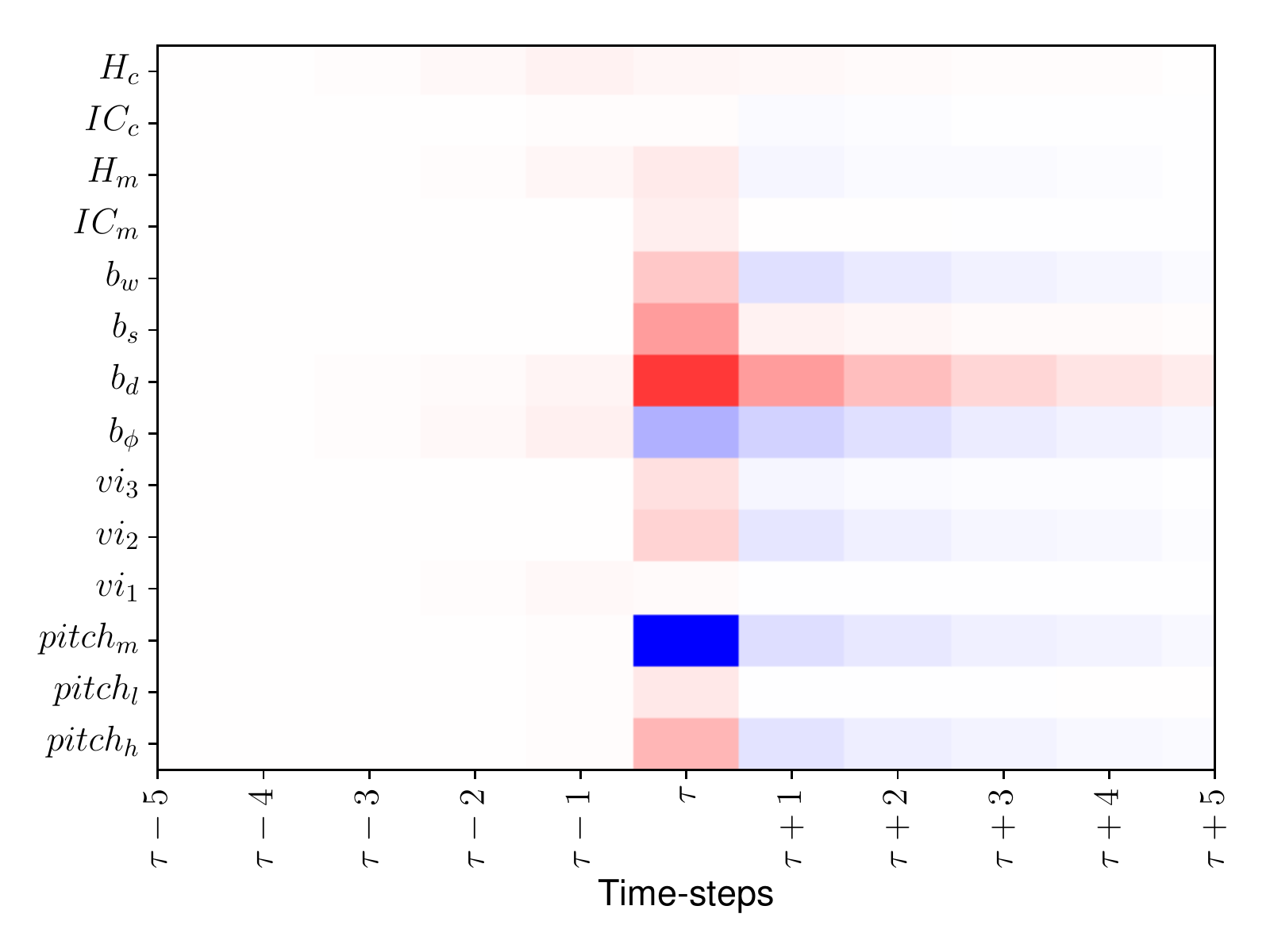}
\includegraphics[trim={10 15 15 10}, clip, width=0.49\linewidth]{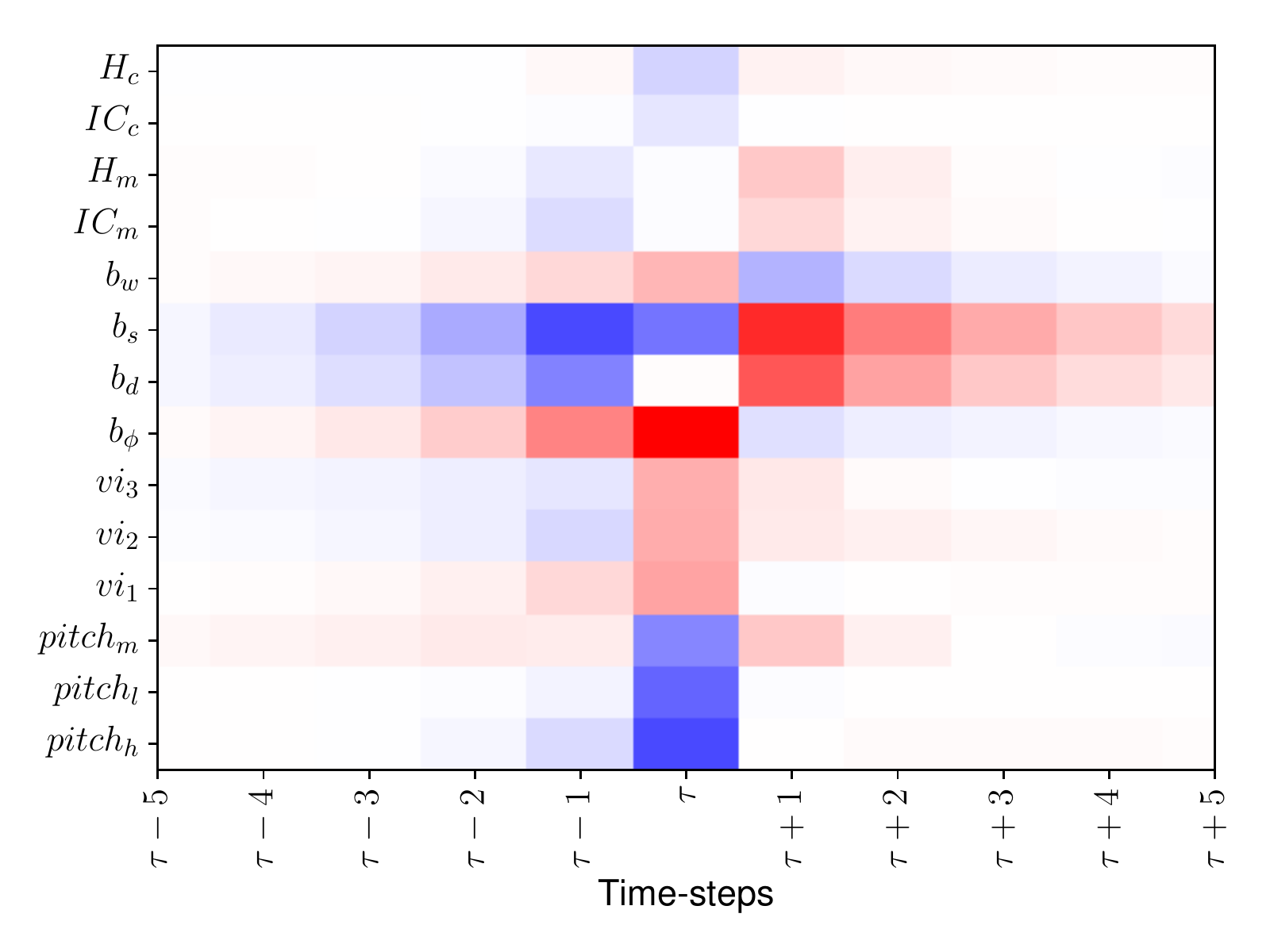}
\caption{Sensitivity plots for $\mathit{BPR}$ (left) and $\mathit{BPR}_d$ (right).
Each row in the plot corresponds to an input feature and each column to the contribution of its value at that time-step to the output of the model at $\tau$ (the center of each plot). Red and blue indicate a positive and negative contribution, respectively.
\vspace{-1.3cm}
 }\label{fig:sensitivity}
\end{center}
\end{figure}

\section{Conclusions}\label{sec:conclusions}
\vskip-1ex
In this paper we presented a model for predicting expressive tempo and dynamics using a combination of expectancy and score features.
Our results support the view that expectancy features, as reflecting what a listener is expecting to hear, can be used to predict the way pianists perform a piece.
The sensitivity analysis also found some evidence relating to well-known rules/guidelines for performance\cc{~\cite{Friberg:2006hs,Gingras:2016bm}.}
\cc{Future work may include the use of expectancy features in combination with larger sets of score descriptors (such as those in \cite{Giraldo:2016iv,CancinoChacon:2017ht}), and derive expectancy features from deep probabilistic models trained directly on (polyphonic) piano-roll representations.}
\vspace{-0.3cm}

\paragraph{Acknowledgements}
  This work has been funded by the European Research Council (ERC) under the EU’s Horizon 2020 Framework Programme (ERC Grant Agreement No. 670035, project CON ESPRESSIONE).
\vspace{-0.3cm}
\bibliographystyle{splncs03}
\bibliography{bib_cc,bib_mg}

\end{document}